\documentclass[12pt,a4paper]{article}
\usepackage{graphicx}
\usepackage{amssymb}
\usepackage{amsmath}
\usepackage{bm}
\usepackage{color}
\usepackage{tikz}
\usepackage{theorem}
\usepackage{cite}
\usepackage{amsfonts}
\usepackage{multirow}

\usepackage{graphicx}
\usepackage{caption}
\usepackage{subcaption}
\usepackage{nicematrix}

\usepackage{geometry}
\usepackage[compat=1.0.0]{tikz-feynman}

\usepackage[sort&compress,numbers, merge]{natbib}

\newcommand{\eprint}[2][]{\href{https://arxiv.org/abs/#2}{arXiv:~\nolinkurl{#2}}}

\definecolor{Orange}{cmyk}{0,0.61,0.87,0}
\definecolor{JungleGreen}{cmyk}{0.99,0,0.52,0} 
\definecolor{OliveGreen}{cmyk}{0.64,0,0.95,0.40}
\definecolor{Brown}{cmyk}{0,0.81,1,0.60}
\definecolor{RoyalBlue}{cmyk}{0.71,0.53,0,0.12}
\definecolor{darkspringgreen}{rgb}{0.09, 0.45, 0.27}

\newcommand{\be}{\begin{equation}}
\newcommand{\ee}{\end{equation}}
\newcommand{\bea}{\begin{eqnarray}}
\newcommand{\eea}{\end{eqnarray}}
\newcommand{\eq}[1]{Eq.~(\ref{#1})}

\newcommand{\Tr}{{\rm Tr\hskip 2pt}}

\usetikzlibrary{decorations.markings}
\usetikzlibrary{fadings}
\usetikzlibrary{shapes.misc}
\tikzset{crossr/.style={cross out, draw=red, minimum size=4*(#1-\pgflinewidth), inner sep=0pt, outer sep=0pt},
crossr/.default={2pt}}
\tikzset{crossb/.style={cross out, draw=black, minimum size=4*(#1-\pgflinewidth), inner sep=0pt, outer sep=0pt},
crossb/.default={2pt}}
\tikzset{crossp/.style={cross out, draw=violet, minimum size=4*(#1-\pgflinewidth), inner sep=0pt, outer sep=0pt},
crossp/.default={2pt}}

\tikzfading %strangely gives bad bounding box when inside the tikzpicture
[
  name=fade out,
  inner color=transparent!0,
  outer color=transparent!100
]

\usepackage[colorlinks=True, citecolor=blue, linkcolor=blue, urlcolor=black]{hyperref}

\allowdisplaybreaks[1]

\renewcommand{\arraystretch}{1.3}

\begin{document}

\begin{titlepage}
\begin{center}
\hfill

\vspace{2.0cm}
{\Large\bf  
%Bootstrapping the Chiral Anomaly
Bootstrapping the Chiral Anomaly at Large $N_c$}

\vspace{2cm}
{\bf 
Teng Ma$^{1,2}$, Alex Pomarol$^{1,3}$ and Francesco Sciotti$^1$
}
\\
\vspace{0.7cm}
{\it\footnotesize
${}^1$IFAE and BIST, Universitat Aut\`onoma de Barcelona, 08193 Bellaterra, Barcelona\\
${}^2$ ICTP-AP, 
%International Centre for Theoretical Physics Asia-Pacific
University of Chinese Academy of Sciences, 100190 Beijing, China\\
${}^3$Departament de F\'isica, Universitat Aut\`onoma de Barcelona, 08193 Bellaterra, Barcelona\\
}

\vspace{0.9cm}
\abstract
The bootstrap approach (demanding  consistency conditions to scattering amplitudes)
has  shown to be quite  powerful  to   tightly constrain  gauge theories at large $N_c$.
We extend previous analysis  to  scattering amplitudes  involving pions and external gauge bosons.
 These  amplitudes  allow us to access  the chiral anomaly 
 and connect low-energy physical quantities to UV properties of the theory.
In particular,  we are able to obtain an analytic
bound  on the chiral anomaly coefficient as a function of the  pion dipole polarizabilities.
This bound can  be  useful for holographic models whose  dual UV completions are not known, and provide  a 
consistency condition to lattice simulations.
\\ \ \\

\end{center}
\end{titlepage}
\setcounter{footnote}{0}

\section{Introduction}

%NOTE: we never discuss the Froissart martin bound \cite{Froissart:1961ux,Martin:1962rt,Jin:1964zza,Bellazzini:2019bzh} \fr{mention it above eq.12} 

In  the large-$N_c$ limit 
strongly-coupled  gauge theories  find a dual description 
as   weakly-coupled theories of mesons and glueballs \cite{tHooft:1973alw,Witten:1979kh}.
This  has allowed to   obtain valuable information on QCD in  the strongly-coupled  regime.
In spite of this,
%the many successful predictions of large-$N_c$ QCD, however,
the fact that the theory contains infinite number of states with all spins
has  made difficult to  get quantitative  predictions on physical quantities. 
 
Recently \cite{Albert:2022oes,Fernandez:2022kzi}, 
it has been shown  that   a  better quantitative understanding  of  large-$N_c$ theories
can be achieved   by requiring consistency conditions, such as unitarity and causality,
to the scattering amplitudes  \cite{Adams:2006sv,Arkani-Hamed:2020blm,deRham:2017avq,Bellazzini:2020cot,Sinha:2020win,Tolley:2020gtv,Caron-Huot:2020cmc,Caron-Huot:2021rmr,Henriksson:2021ymi}.
Thanks to the  simple
analytical structure of the scattering amplitudes at   large-$N_c$,
positivity constraints have provided precise  and strong bounds on low-energy quantities, such as 
Wilson coefficients of the  chiral Lagrangian as well as meson couplings.
High-spin states  contributions to these low-energy physical quantities  have shown to be small, giving 
an understanding of the  phenomenological successes of Vector Meson Dominance and   holographic QCD
\cite{Fernandez:2022kzi}. 

We extend these previous analysis  to  scattering amplitudes of Goldstones ($\pi$ and $\eta$) interacting with external gauge bosons ($W$).
The main purpose is to access  information related to the chiral anomaly and other physical quantities such as 
 pion polarizabilities.
In particular, we will analyze  the amplitudes $W\pi\to W\pi$ and $W\pi\to \eta\pi$,
this later being proportional, at low-energies,  to the chiral anomaly.
We will understand the  meson exchange selection rules 
that apply to these  amplitudes  and that will help us to derive the sign of their on-shell residues.
By demanding a good high-energy behaviour of the scattering amplitudes, we will derive  dispersion relations that 
will allow us to obtain interesting bounds on  low-energy quantities.\footnote{Previous studies in the context of  QCD can be found in \cite{martingourdin,Ko:1989yd,Gasser:2005ud,Filkov:2005ccw,Gasser:2006qa,Burgi:1996qi,Dai:2016ytz,Ren:2022fhp,Lee:2023rmz}.}

Our main result will be the derivation of an upper bound on the coefficient of the chiral anomaly~$\kappa$:
\be
\frac{\kappa}{\sqrt{{\cal P}/F^2_\pi}} \leq \sqrt{\frac{1}{2}}\,,
\ee
where $F_\pi$ is the pion decay constant, and  ${\cal P}$, to be specified later,  is related to   pion  polarizabilities. 
This  bound provides a non-trivial  
connection between  low-energy physical quantities and  UV properties of the theories.\footnote{Bounds on the $a$-anomaly 
from scattering amplitudes  requirements have been derived in \cite{Karateev:2022jdb,Marucha:2023vrn}.}
On the other hand,  for  phenomenological models for QCD, such as a holographic or NJL models,
the above upper bound can provide  a valuable  consistency check, as well as for  lattice simulations.

The analysis described here can also be useful to derive constraints on a large number of  low-energy physical constants,
specially to those related to the electromagnetic properties of pions and other mesons.

The  paper is organized as follows. In Sec.~\ref{sec:amplitude} we discuss the properties of amplitudes for pions and gauge bosons and derive the dispersion relations and null constraints. In Sec.~\ref{sec:bound}, we analytically derive an upper bound on the chiral anomaly coefficient. We conclude in~Sec. \ref{sec:conclusion}. The appendices contain details about  the classification of mesons and   the signs of residues (\ref{appa}), pion polarizabilities (\ref{pola}),
 numerical bounds on Wilson coefficients (\ref{wilsonsapp}),
 and  $su$-models (\ref{sumodelapp}).

\section{Amplitudes for pions and external gauge bosons in the large-$N_c$ limit}
\label{sec:amplitude}
We are interested in studying $SU(N_c)$ gauge theories in the large-$N_c$ limit.
We will consider that the  theory also contains $N_f$ massless quarks ($q_L,q_R$) in the fundamental representation of $SU(N_c)$,   with the following pattern of chiral symmetry breaking: $U(N_f)_L\times U(N_f)_R\rightarrow SU(N_f)\times U(1)$. The massless Goldstone bosons associated with the breaking of this global symmetry are the pions ($\pi^a$ with $a=1,...,N_f^2-1$ in the adjoint representation of $SU(N_f)$) and the $\eta$ (singlet). 
The chiral axial symmetry $U(1)_A$ is anomalous but in the large-$N_c$ limit
the corresponding Goldstone, the $\eta$, remains massless.

As usual, we  will
introduce the coupling to external gauge bosons by gauging the global
$SU(N_f)\times U(1)$ symmetry.   These gauge bosons will be considered   (non-propagating)  external fields,  sitting in the adjoint and singlet representations.

For simplicity, we will concentrate in this paper in  the  $N_f=2$  case, but the arguments we make can be extended straightforwardly to a general $N_f$. For $N_f=2$  the pions and gauge bosons are   isospin triplets, $\pi^a$ and $W^a$, and  isospin singlets, the $\eta$ and  $B$.\footnote{We normalize the gauge bosons as $W^aT^a+B\, \mathbb{I}/2$, where 
$\Tr[T^aT^b]=1/2$, and the gauge coupling $g=1$.} 
%The scattering amplitudes we will consider will have the particles introduced above as asymptotic states, in particular we will study the following three classes of amplitudes
We are interested in the following amplitudes:
\begin{eqnarray}
 \begin{tikzpicture}[baseline=.1ex]
  \begin{feynman}[every blob={/tikz/fill=blue!30}]
    \vertex[blob] (m) at (0, 0){};
    \vertex (a) at (-1,-1) {$\pi^b$} ;
    \vertex (b) at ( 1,-1) {$\pi^c$};
    \vertex (c) at (-1, 1) {\small{$W^a$}};
    \vertex (d) at ( 1, 1) {$\eta$};
    \vertex (e) at (-0.4,0.8) {\scriptsize{1}} ;
    \vertex (e) at (-0.4,-0.8) {\scriptsize{2}} ;
    \vertex (e) at (0.4,0.8) {\scriptsize{3}} ;
    \vertex (e) at (0.4,-0.8) {\scriptsize{4}} ;
    \diagram* {
      (a) -- [dashed] (m) -- [photon] (c),
      (b) -- [dashed] (m) -- [dashed] (d),
    };
  \end{feynman}
\end{tikzpicture}\qquad\quad&\qquad\quad
\begin{tikzpicture}[baseline=.1ex]
  \begin{feynman}[every blob={/tikz/fill=blue!30}]
    \vertex[blob] (m) at (0, 0){};
    \vertex (a) at (-1,-1) {$\pi^b$} ;
    \vertex (b) at ( 1,-1) {$\pi^d$};
    \vertex (c) at (-1, 1) {\small{$W^a$}};
    \vertex (d) at ( 1, 1) {\small{$W^c$}};
    \vertex (e) at (-0.4,0.8) {\scriptsize{1}} ;
    \vertex (e) at (-0.4,-0.8) {\scriptsize{2}} ;
    \vertex (e) at (0.4,0.8) {\scriptsize{3}} ;
    \vertex (e) at (0.4,-0.8) {\scriptsize{4}} ;
    \diagram* {
      (a) -- [dashed] (m) -- [photon] (c),
      (b) -- [dashed] (m) -- [photon] (d),
    };
  \end{feynman}
\end{tikzpicture}\qquad\quad&\qquad\quad
\begin{tikzpicture}[baseline=.1ex]
  \begin{feynman}[every blob={/tikz/fill=blue!30}]
    \vertex[blob] (m) at (0, 0){};
    \vertex (a) at (-1,-1) {$\pi^b$} ;
    \vertex (e) at (-0.4,0.8) {\scriptsize{1}} ;
    \vertex (e) at (-0.4,-0.8) {\scriptsize{2}} ;
    \vertex (e) at (0.4,0.8) {\scriptsize{3}} ;
    \vertex (e) at (0.4,-0.8) {\scriptsize{4}} ;
    \vertex (b) at ( 1,-1) {$\pi^d$};
    \vertex (c) at (-1, 1) {$\pi^a$};
    \vertex (d) at ( 1, 1) {$\pi^c$};
    \diagram* {
      (a) -- [dashed] (m) -- [dashed] (c),
      (b) -- [dashed] (m) -- [dashed] (d),
    };
  \end{feynman}
  \end{tikzpicture} \label{amplitudes} \\
(a)\qquad \qquad \quad \  & (b) &\qquad \qquad \quad \ \  (c)
\nonumber
\end{eqnarray}
The amplitude (a) is important as, at low-energies,  becomes proportional to the chiral anomaly coefficient that we want to bootstrap. For this purpose,  as we will see, we will also need  the amplitudes (b) and (c).
The amplitude (c) was already discussed  in detail in \cite{Albert:2022oes,Fernandez:2022kzi}.
% and therefore throughout the paper we will only recall the necessary definitions of Wilson coefficients as a function of the couplings. 
In this section we will then only discuss the amplitudes (b) and (a) in the large-$N_c$ limit for the different 
% We begin giving a general overview of the $\pi W\rightarrow \pi W$ amplitude (b) for the two possible 
 polarizations of the gauge bosons $\lambda_{W}=\pm1$.
 %, either for equal polarizations  for the elastic and the inelastic process.  
 In the following, we will derive the relevant  dispersion relations necessary to  obtain sum rules for the  chiral anomaly coefficient and other Wilson coefficients, as well as  null constraints necessary for obtaining  bounds.

\subsection{Exchanged meson states}

%Before studying in detail the amplitudes, we must initially understand what states can be exchanged in the various channels. 
In the large-$N_c$ limit, a $SU(N_c)$ gauge theory  reduces to a theory of weakly-coupled mesons,  whose trilinear couplings scale as $\sim 1/\sqrt{N_c}$  \cite{tHooft:1973alw,Witten:1979kh}. 
Amplitudes are then mediated by tree-level  meson exchange. 
In the amplitudes  (\ref{amplitudes}), the exchanged mesons are colorless  $q\bar q$ states that, as in the quark model, can be classified according to their  quantum numbers: 
 Isospin ($I$),  $G-$parity ($G$), parity ($P$) and spin ($J$).
The relation between these quantum numbers is discussed in   Appendix~\ref{appa}.\footnote{Fixing $I$, $G$ and $P$,  the spin $J$ is  fixed to either be  even or odd, as explicitly shown in Table~\ref{tab:states}.}
 We have six types of mesons  as shown in Table~\ref{tab:states}.
It is important to say that in the large-$N_c$ limit, each $I=1$ state (left column) has an  $I=0$ associated state
(the one next in the right column) of equal mass and related couplings due to the underlying $U(2)$ symmetry \cite{tHooft:1973alw,Witten:1979kh}.

\begin{table}[!]
\begin{center}
\renewcommand{\arraystretch}{1.6}
\begin{NiceTabular}[corners,hvlines]{|c|c||c|c|c|}
\CodeBefore
\cellcolor{yellow!20}{1-2,1-3}
  \cellcolor{blue!20}{2-1,3-1,4-1,5-1,6-1,7-1,2-4,3-4,4-4,5-4,6-4,7-4}
  \cellcolor{green!20}{3-2,3-3,6-3,6-2}
\Body
 & Isospin $I=0$  & Isospin $I=1$ &  & $n$  \\
 \hline
\Block{3-1}{$G=+1$} & $J_\text{odd}^+$ ($f_J$)  & $J_\text{odd}^+$  ($a_J$)  & \Block{3-1}{$G=-1$}&  1 \\
                 & $J_\text{even}^+$      ($f_J$)          & $J_\text{even}^+  $    ($a_J$)  &        &   2 \\
                  & $J_\text{even}^- $       ($\eta_J$)         & $J_\text{even}^-  $      ($\pi_J$)  &       &   3  \\
                  \hline
\Block{3-1}{$G=-1$} & $J_\text{odd}^+$  ($h_J$)  & $J_\text{odd}^+$ ($b_J$)& \Block{3-1}{$G=+1$}&   4\\
                 & $J_\text{odd}^-$          ($\omega_J$)      & $J_\text{odd}^-  $   ($\rho_J$)  &      &   5 \\
                  & $J_\text{even}^- $        ($\omega_J$)        & $J_\text{even}^-  $   ($\rho_J$)  &      &  6  \\

\end{NiceTabular}
\end{center}
\caption{\it List of $q\bar{q}$ states classified in terms of their Isospin, $G-$parity and $J^P$.  We also 
provide the names as they are generally referred to in QCD \cite{ParticleDataGroup:2022pth}.
In green we highlight the states that enter in the amplitude $W\pi\rightarrow\eta\pi$ which is related  to the chiral anomaly.}
\label{tab:states}
\end{table}

\subsection{The $\pi W \rightarrow \pi W $ Amplitudes}
Let us start with  the amplitude (b). 
%The most general way of writing an amplitude whose external states all transform in the adjoint representation of $SU(N_f)$ is 
%\bea
%{\cal M}_{a,b,c,d}= A_s\left[\frac{2}{N_f}\delta_s+d_s\right]
%+A_t\left[\frac{2}{N_f}\delta_t+d_t\right]
%+A_u\left[\frac{2}{N_f}\delta_u +d_u\right]
%+B_s\ \delta_s
%+B_t\ \delta_t
%+B_u\ \delta_u\ ,
%\label{generalsunf}
%\eea
%where $s=(p_a+p_b)^2$, $t=(p_a-p_c)^2$, $u=(p_a-p_d)^2$, and 
%\bea
%&& \delta_s=\delta_{ab}\delta_{cd}\ ,\ \delta_t=\delta_s(b\leftrightarrow c)\ , \ \delta_u=\delta_s(b\leftrightarrow d)\,, \\ \nonumber
%&&d_s=d_{abe}d_{cde}\ , \ d_t=d_s(b\leftrightarrow c)\ , \ d_u=d_s(b\leftrightarrow d)\,,
%\eea
%correspond to the various ways of contracting $SU(N_f)$ adjoint indices into singlets. In the large-$N_c$ limit the $B$ functions vanish as they are associated to double trace operators which are subleading. 
For $N_f=2$, the most general structure we can write for this amplitude takes the form
\be
{\cal M}(W^a,\pi^b,W^c,\pi^d)= A_t(s,u) \delta_t + \tilde{A}(s,u) \delta_s + \tilde{A}(u,s) \delta_u\,,
\label{mainamp}
\ee
where $s=(p_a+p_b)^2$, $t=(p_a-p_c)^2$, $u=(p_a-p_d)^2$, and 
$\delta_s=\delta_{ab}\delta_{cd}\ ,\ \delta_t=\delta_{ac}\delta_{bd}\ , \ \delta_u=\delta_{ad}\delta_{cb}$.
We have demanded the amplitude to be invariant under the  exchange of the two pions
that also requires $A_t(s,u)$ to be $s\leftrightarrow u$ symmetric. 

In the large-$N_c$ limit this amplitude is mediated by  tree-level exchanges of mesons, either  in the $s-$channel, $t-$channel  and/or $u-$channel.  Since $q\bar q$ states cannot have $I=2$, it will be important to focus  in  $I=2$ eigen-amplitudes.
These are given by 
\be
M^{I=2}_s=A_t(s,u)+\tilde{A}(u,s)\,,  \qquad M^{I=2}_t=\tilde{A}(s,u)+\tilde{A}(u,s)\,, \qquad M^{I=2}_u=A_t(s,u)+\tilde{A}(s,u)\,,
\label{isospin2}
\ee 
respectively for the $s$, $t$ and $u-$channel.   These amplitudes  cannot have mesons exchanging in  the   $s$, $t$ and $u-$channel respectively.

In the amplitude \eq{mainamp}  the $s-$channel and $u-$channel    exchanged meson $i$ couples to $W\pi$, while the
$t-$channel  exchanged meson couples to  $WW$ and $W\pi$: 
\be
\begin{tikzpicture}[baseline=.1ex]
  \begin{feynman}[every dot={/tikz/fill=black!70}]
    \vertex[dot] (m1) at (-0.4, 0){};
    \vertex[dot] (m2) at (0.4, 0){};
    \vertex (a) at (-1,-1) {$\pi$} ;
    \vertex (b) at ( 1,-1) {$\pi$};
    \vertex (c) at (-1, 1) {\small{$W$}};
    \vertex (d) at ( 1, 1) {\small{$W$}};
    \diagram* {
      (a) -- [dashed] (m1) -- [photon] (c),
      (m2) -- [double] (m1),
      (b) -- [dashed] (m2) -- [photon] (d),
    };
  \end{feynman}
  \node[text width=2.2cm] at (2.5,0) {$\propto  g_{W\pi i}^2$};
  \end{tikzpicture}
    \begin{tikzpicture}[baseline=.1ex]
  \begin{feynman}[every dot={/tikz/fill=black!70}]
    \vertex[dot] (m1) at (0, -0.4){};
    \vertex[dot] (m2) at (0, 0.4){};
    \vertex (a) at (-1,-1) {$\pi$} ;
    \vertex (b) at ( 1,-1) {$\pi$};
    \vertex (c) at (-1, 1) {\small{$W$}};
    \vertex (d) at ( 1, 1) {\small{$W$}};
    \diagram* {
      (a) -- [dashed] (m1) -- [photon] (d),
      (m2) -- [double] (m1),
      (c) -- [photon] (m2) -- [dashed] (b),
    };
  \end{feynman}
  \node[text width=2.2cm] at (2.5,0) {$\propto  g_{W\pi i}^2$};
\end{tikzpicture}
  \begin{tikzpicture}[baseline=.1ex]
  \begin{feynman}[every dot={/tikz/fill=black!70}]
    \vertex[dot] (m1) at (0, -0.4){};
    \vertex[dot] (m2) at (0, 0.4){};
    \vertex (a) at (-1,-1) {$\pi$} ;
    \vertex (b) at ( 1,-1) {$\pi$};
    \vertex (c) at (-1, 1) {\small{$W$}};
    \vertex (d) at ( 1, 1) {\small{$W$}};
    \diagram* {
      (a) -- [dashed] (m1) -- [dashed] (b),
      (m2) -- [double] (m1),
      (c) -- [photon] (m2) -- [photon] (d),
    };
  \end{feynman}
  \node[text width=2.5cm] at (2.5,0) {$\propto  g_{WWi}g_{\pi\pi i}$};
\end{tikzpicture} 
\label{tpoles}
\ee
Since the couplings $g_{WWi}$ do not appear in the amplitude $W\pi\eta\pi$ (the one related to the chiral anomaly), we will restrict to situations in which  the $t-$channel meson exchange is not present such that  we will not have to deal with
these couplings $g_{WWi}$ any longer.
This  can be achieved by  either taking  $t-$fixed  or  working with $M^{I=2}_t$.
This leaves only 3 possibilities to consider:
\begin{enumerate}
\item
   $M^{I=2}_t$ at $u-$fixed\,, 
   \label{poss2}
\item
  $M^{I=2}_t$ at $t-$fixed\,,
  \label{poss1}
\item
    $M^{I=2}_u$ at $t-$fixed\,.
    \label{poss3}
   \end{enumerate}
Notice that we have not included   $M^{I=2}_s$ since it contains the same information as $M^{I=2}_u$ as far as our positivity arguments are concerned. 

%The coupling relevant to the chiral anomaly is the first, therefore throughout the paper we will ignore all channels involving the latter coupling. The second coupling becomes  relevant if one wants to connect the 4 Pion amplitude to the 4 Photon amplitude, which might reveal interesting structures between the Wilsons but is not the focus of this work.

%If one wants to impose positivity bounds, 
It is also of central importance to understand the sign of the on-shell couplings of the meson being exchanged, or, equivalently, the sign of the residues at the mass poles. Defining these as $R_{I=1,0}$ for the 
$I=1,0$ mesons respectively, we have 
%Initially let us notice that by choosing particular configurations of isospin of the external particles it is possible to select wither the function $A_t$ or $\tilde{A}$ and obtain important insights on the isospin of the resonance being exchanged in a channel. To give a few examples the on-shell amplitude in the $s-$channel becomes (keeping the helicity structure general)
\begin{eqnarray*}
\begin{tikzpicture}
  \node[text width=7cm] at (-5,0) {${\cal M}(\delta_t\neq0,\delta_s=\delta_u=0) = A_t(s,u)\   \longrightarrow $};
  \node[text width=7cm] at (1.1,-0.3) {\tiny{$s-$channel}};
  \node[text width=7cm] at (1.3,-0.6) {\tiny{onshell}};
  \begin{feynman}[every dot={/tikz/fill=black!70}]
    \vertex[dot] (m1) at (-0.4, 0){};
    \vertex[dot] (m2) at (0.4, 0){};
    \vertex (a) at (-1,-1) {$\pi^a$} ;
    \vertex (b) at ( 1,-1) {$\pi^a$};
    \vertex (c) at (-1, 1) {\small{$W^b$}};
    \vertex (d) at ( 1, 1) {\small{$W^b$}};
    \diagram* {
      (a) -- [dashed] (m1) -- [photon] (c),
      (m2) -- [double] (m1),
      (b) -- [dashed] (m2) -- [photon] (d),
    };
  \end{feynman}
  \node[text width=4cm] at (4,0) {\large{$=- \sum_i  \frac{(R_{I=1})_i}{s-m_i^2}$},};
\end{tikzpicture} \\
\begin{tikzpicture}
  \node[text width=7cm] at (-4.9,0) {${\cal M}(\delta_s\neq0,\delta_t=\delta_u=0) = \tilde{A}(s,u)\    \longrightarrow $};
      \node[text width=7cm] at (1.1,-0.3) {\tiny{$s-$channel}};
  \node[text width=7cm] at (1.3,-0.6) {\tiny{onshell}};
  \begin{feynman}[every dot={/tikz/fill=black!70}]
     \vertex[dot] (m1) at (-0.4, 0){};
    \vertex[dot] (m2) at (0.4, 0){};
    \vertex (a) at (-1,-1) {$\pi^b$} ;
    \vertex (b) at ( 1,-1) {$\pi^a$};
    \vertex (c) at (-1, 1) {\small{$W^b$}};
    \vertex (d) at ( 1, 1) {\small{$W^a$}};
    \diagram* {
      (a) -- [dashed] (m1) -- [photon] (c),
      (m2) -- [double] (m1),
      (b) -- [dashed] (m2) -- [photon] (d),
    };
  \end{feynman}
  \node[text width=4cm] at (4,0)  {\large{$= -\sum_i  \frac{(R_{I=0})_i}{s-m_i^2}$},};
\end{tikzpicture} 
\end{eqnarray*}
where the sum over $i$ runs over all the possible intermediate states. Knowing that the amplitudes 
\eq{isospin2} cannot have poles at the $s$, $t$ and $u$--channel respectively, we can derive 
 the relative sign of the residues in the $s$ and $u-$channel, which we show in Table~\ref{tab:tablecouplings}.
\begin{table}[h!]
\begin{center}
\begin{tabular}{ |p{4cm}|| c | c | c |  }
 \hline
\centering{Meson channel}& $-A_t(s,u)$ & $-\tilde{A}(s,u)$ & $-\tilde{A}(u,s)$ \\[0.5ex] 
 \hline
\centering{$s-$channel}  & $ R_{I=1}$  & $R_{I=0}$   &  $-R_{I=1}$\\
\centering{$u-$channel} & $ R_{I=1}$   & $ -R_{I=1}$ & $R_{I=0}$\\
 \hline
\end{tabular}
\caption{\it Residues  in the $s$ and $u-$channel of the amplitudes $A_t$ and $\tilde{A}$.}
\label{tab:tablecouplings}
\end{center}
\end{table}

For  $I=1\, (0)$ mesons, we have 3 types of states, $n=1,2,3\, (4,5,6)$,  given in  Table~\ref{tab:states}.
The signs of their residues  depend  on the helicity of the external vector states. 
 We   have  two independent possibilities,
 either an elastic process ($W^+ \pi\to W^+ \pi$) which we will refer to as the $+-$ amplitude, 
 or inelastic ($W^+ \pi\to W^- \pi$) which we will refer to as the $++$ amplitude.
Following the discussion of  Appendix~\ref{appa},  one finds that the signs of the residues are given by 
\bea
R_{I=0}^{+-} &=&+g_4^2+g_5^2+g_6^2 \qquad \qquad R_{I=1}^{+-} = +g_1^2+g_2^2+g_3^2\,,\nonumber \\
R_{I=0}^{++} &=& -g_4^2+g_5^2-g_6^2  \qquad \qquad R_{I=1}^{++} = -g_1^2+g_2^2-g_3^2\,,
\label{Rstates}
\eea
where  we have defined
\be
g_n^2\equiv g_{\pi W,n}^2>0\ ,\ \ \  n=1,...,6\,,
\label{defgn}
\ee
 and $n$ labels the different type of mesons as defined in Table~\ref{tab:states}.

The helicity configuration forces a particular kinematic pre-factor in the amplitude which, in the spinor helicity formalism, can be written as 
\be
{\cal M}^{+-} \propto \langle12\rangle^2 [23]^2 \propto su  \quad ,\quad  {\cal M}^{++} \propto [13]^2\propto s+u \,,
\ee
where the numbers label the states as in Fig.~\ref{amplitudes} (b).

At energies below the mass $M$ of the lightest massive meson,  we can   expand  the amplitudes
in powers of $s/M,u/M\to 0$. 
For the elastic scattering amplitude we have 
\bea
M_t^{I=2}(s,u)|_{+-}
&=&2+s u\bigg( \sum_{m=0}^\infty\sum_{l=0}^{[m/2]} \, g_{m,l}\,  s^{\{m-l}  u^{l\}}\bigg)\label{elastict}\\
&=& 2+s u \big(g_{0,0}+ g_{1,0}\,(s+u)+g_{2,0}\,(s^2+u^2)+ 2g_{2,1} su + ...) \ \nonumber,\\
M_{u}^{I=2}(s,u)|_{+-}
&=&-1 - s u\bigg( \sum_{m=0}^\infty\sum_{l=0}^{[m/2]} \, h^s_{m,l}\,  s^{\{m-l}  u^{l\}}+\sum_{m=0}^\infty\sum_{l=0}^{[m/2]} \, h^a_{m,l}\,  s^{[m-l}  u^{l]}\bigg)\label{elasticu}\\
&=& -1- s u \big(h^s_{0,0}+ h^s_{1,0}\,(s+u)+h^a_{1,0}\,(s-u)+h^s_{2,0}\,(s^2+u^2)+...)\ \nonumber,
\eea
 where \eq{elastict}  is fully $s\leftrightarrow u$ symmetric, and in \eq{elasticu} we have  separated the symmetric and the antisymmetric part.  The first terms of the expansion correspond to the pion exchange whose gauge coupling we normalize to $1$.
On the other hand, for the inelastic scattering amplitude we have 
\bea
M_t^{I=2}(s,u)|_{++}
&=&(s+u) \bigg( \sum_{m=0}^\infty\sum_{l=0}^{[m/2]} \, f_{m,l}\,  s^{\{m-l}  u^{l\}}\bigg)\label{inelastict}
\\
&=&(s+u) \big(f_{0,0}+ f_{1,0}\,(s+u)+f_{2,0}\,(s^2+u^2)+ f_{2,1} su+...)\ \nonumber\,.
\eea
We do not need to give  the expression for the low-energy expansion of $M_u^{I=2}(s,u)|_{++}$ since, as we will see, this amplitude does not provide  any extra information for bounding the  chiral anomaly. The low-energy constants
$g_{m,l}$, $h^s_{m,l}$, $h^a_{m,l}$ and $f_{m,l}$, referred here as Wilson coefficients, are the low-energy parameters that we want to constrain. 

\subsection{Dispersion Relations and Null Constraints}

%The analytic structure for the amplitudes in \ref{poss1}-\ref{poss3}   is shown in Fig.~\ref{fig:sumrules}.
In order to derive  sum rules and null constraints from our dispersion relations, 
we will assume that for all $k\geq{k_{\rm min}}$, the amplitudes behave as follows at high energies ($k_\text{min}$ will be indicated for the various processes):\\
%\vspace{12mm}
\begin{subequations}\label{UVeq}
\begin{minipage}[t]{0.4\linewidth}
\begin{equation}
\lim_{|s|\to\infty}\frac{M_t^{I=2}(s,u)}{s^k} \to 0\,, \label{UVa}
\end{equation}
  \end{minipage}%
  \hspace{0.1\textwidth}
  \begin{minipage}[t]{0.4\linewidth}
  \vspace{-4mm}
\begin{equation}
\lim_{|s|\to\infty}\frac{M_t^{I=2}(s,-t-s)}{s^k} \to 0\,, \label{UVb}
\end{equation}
    \end{minipage}\\
    \vspace{-6mm}
    \begin{center}
    \begin{minipage}{0.4\textwidth}
\begin{equation}
\lim_{|s|\to\infty}\frac{M_u^{I=2}(s,-s-t)}{s^k} \to 0\,. \label{UVc}
\end{equation}
  \end{minipage}%
  \end{center}
\end{subequations}
%\fr{Generally, the consistent amplitude must satisfy the  Froissart’s bound, i.e. $ \lim_{s \to \infty}M^{I=2}_t(s,u)/u^2 \to 0 $ for $u$ fixed, so $k_{\text{min}} \le 2$.} 
 The amplitudes can be expanded in partial waves in the physical regions.  For the elastic process we have
\bea
\label{pwpm}
{\rm Im} M_{u,t}^{I=2}(s,u)\big|_{+-} &=& \sum_i (2J+1)\rho^{+- ,\pm}_J(s) d_{1,1}^J\big(\cos \theta_s \big)\,,\nonumber \\
(2J+1)\rho^{+-,\pm}_J(s) &=&\pi\sum_i \bigg( R_{I=0}^{+-}\pm R_{I=1}^{+-}\bigg)_i m_i^2 \delta(s-m_i^2) \delta_{J J_i}\,,
\eea
where the $u,t$ subscript refers to the $\pm$ spectral density
and  $\cos\theta_s=(t-u)/s$. For the inelastic amplitude similarly we have 
\bea
\label{pwpp}
{\rm Im} M_{u,t}^{I=2}(s,u)\big|_{++} &=& \sum_i (2J+1)\rho^{++ ,\pm}_J(s) d_{1,-1}^J\big(\cos \theta_s \big)\,, \nonumber\\
(2J+1)\rho^{++,\pm}_J(s) &=&\pi\sum_i \bigg(R_{I=0}^{++}\pm R_{I=1}^{++}\bigg)_i m_i^2 \delta(s-m_i^2) \delta_{J J_i}\,.
\eea

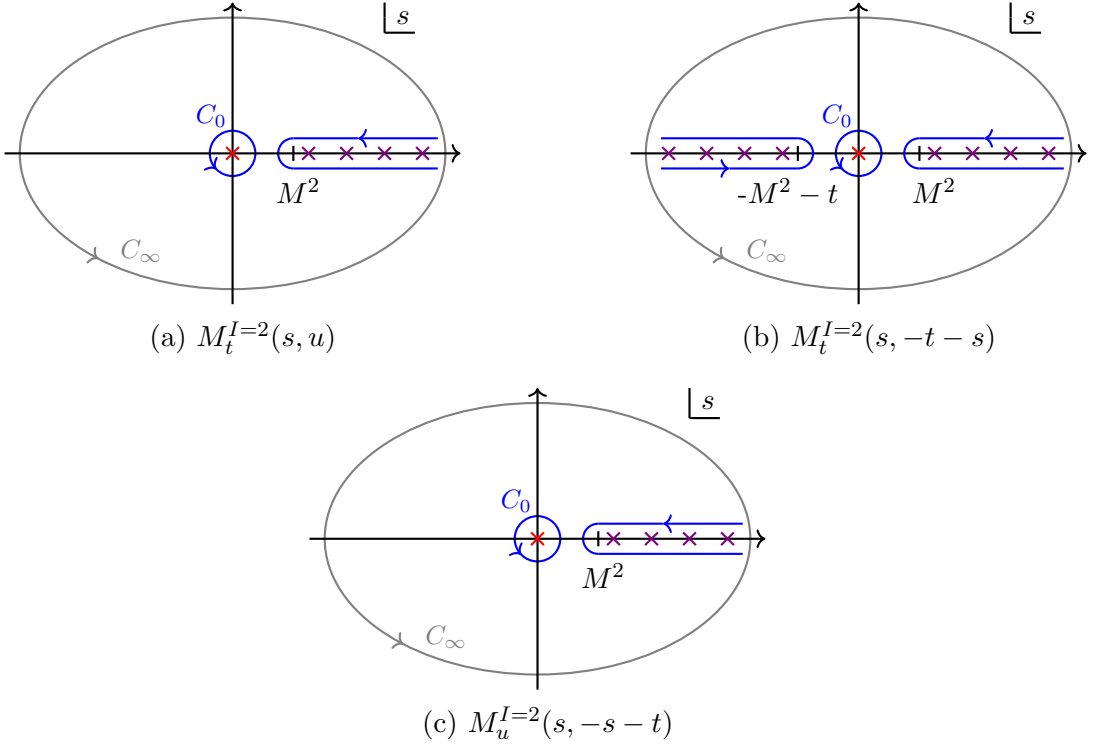
\begin{figure}[t]\hspace{0.1cm}
  \begin{subfigure}[b]{0.5\textwidth}

\begin{tikzpicture}
\begin{scope}[thick,font=\small]
\hskip1.1cm 
\draw[decoration={markings, mark=at position 0.625 with {\arrow{>}}},postaction={decorate}, color=gray] (3,0) ellipse (2.8 and 1.8) {};
\draw[decoration={markings, mark=at position 0.625 with {\arrow{>}}},postaction={decorate}, color=blue] (3,0) circle (0.3) {};
\node[color=gray] at (1.8,-1.3) {\footnotesize{$C_\infty$}};
\node[color=blue] at (2.72,0.5) {\footnotesize{$C_0$}};
\draw[color=blue] (3.8,0.2)--(4.65,0.2) {};
\draw[color=blue,<-] (4.65,0.2)--(5.7,0.2) {};
\draw[color=blue] (3.8,-0.2)--(5.7,-0.2) {};
\draw[color=blue] (3.8,0.2) arc (90:270:0.2) {};
\draw[black, thick,->] (0,0) -- (6,0);
\draw[black, thick,->] (3,-2) -- (3,2);
\draw[black] (5,2) -- (5,1.8) node[right] {$s$};
\draw[black] (5,1.8) -- (5,1.6) {};
\draw[black] (5,1.6) -- (5.4,1.6) {};
\draw (3,0) node[crossr] {};
\draw (4,0) node[crossp] {};
\draw[black]  (3.8,0.1) -- (3.8,-0.1) node[midway,below=0.18cm,align=center] {$\ M^2$};
\draw (4.5,0) node[crossp] {};
\draw (5,0) node[crossp] {};
\draw (5.5,0) node[crossp] {};
\end{scope}
\end{tikzpicture}
     \caption{
  $M_t^{I=2}(s,u)$ 
    % pole structure for $u<0.\qquad\qquad\qquad$
  }
    \label{fig:uchannel}
  \end{subfigure}\hspace{-0.4cm}
  \begin{subfigure}[b]{0.5\textwidth}
\begin{tikzpicture}
\begin{scope}[thick,font=\small]
\hskip1.1cm 
\draw[decoration={markings, mark=at position 0.625 with {\arrow{>}}},postaction={decorate}, color=gray] (3,0) ellipse (2.8 and 1.8) {};
\draw[decoration={markings, mark=at position 0.625 with {\arrow{>}}},postaction={decorate}, color=blue] (3,0) circle (0.3) {};
\node[color=gray] at (1.8,-1.3) {\footnotesize{$C_\infty$}};
\node[color=blue] at (2.72,0.5) {\footnotesize{$C_0$}};
\draw[black] (5,2) -- (5,1.8) node[right] {$s$};
\draw[black] (5,1.8) -- (5,1.6) {};
\draw[black] (5,1.6) -- (5.4,1.6) {};
\draw[black, thick,->] (0,0) -- (6,0);
\draw[black, thick,->] (3,-2) -- (3,2);
\draw (3,0) node[crossr] {};
\draw (4,0) node[crossp] {};
\draw[black]  (3.8,0.1) -- (3.8,-0.1) node[midway,below=0.18cm,align=center] {$\ \ \  M^2$};
\draw[black]  (2.2,0.1) -- (2.2,-0.1) node[midway,below=0.18cm,align=center] {-$M^2-t \ \ $};
\draw (4.5,0) node[crossp] {};
\draw (5,0) node[crossp] {};
\draw (5.5,0) node[crossp] {};
\draw (0.5,0) node[crossp] {};
\draw (1,0) node[crossp] {};
\draw (1.5,0) node[crossp] {};
\draw (2,0) node[crossp] {};
\draw[color=blue] (3.8,0.2)--(4.65,0.2) {};
\draw[color=blue,<-] (4.65,0.2)--(5.7,0.2) {};
\draw[color=blue] (3.8,-0.2)--(5.7,-0.2) {};
\draw[color=blue] (3.8,0.2) arc (90:270:0.2) {};
\draw[color=blue] (0.4,0.2)--(2.2,0.2) {};
\draw[color=blue,->] (0.4,-0.2)--(1.3,-0.2) {};
\draw[color=blue] (1.3,-0.2)--(2.2,-0.2) {};
\draw[color=blue] (2.2,0.2) arc (90:-90:0.2) {};
\end{scope}
\end{tikzpicture}
    \caption{$M_{t}^{I=2}(s,-t-s)$ 
      %pole structure for $u<0.\qquad\qquad$
    }
    \label{fig:tchannel}
  \end{subfigure}
  \begin{center}
\begin{subfigure}[c]{0.5\textwidth}
\begin{tikzpicture}
\begin{scope}[thick,font=\small]
\hskip1.1cm 
\draw[decoration={markings, mark=at position 0.625 with {\arrow{>}}},postaction={decorate}, color=gray] (3,0) ellipse (2.8 and 1.8) {};
\draw[decoration={markings, mark=at position 0.625 with {\arrow{>}}},postaction={decorate}, color=blue] (3,0) circle (0.3) {};
\node[color=gray] at (1.8,-1.3) {\footnotesize{$C_\infty$}};
\node[color=blue] at (2.72,0.5) {\footnotesize{$C_0$}};
\draw[color=blue] (3.8,0.2)--(4.65,0.2) {};
\draw[color=blue,<-] (4.65,0.2)--(5.7,0.2) {};
\draw[color=blue] (3.8,-0.2)--(5.7,-0.2) {};
\draw[color=blue] (3.8,0.2) arc (90:270:0.2) {};
\draw[black, thick,->] (0,0) -- (6,0);
\draw[black, thick,->] (3,-2) -- (3,2);
\draw[black] (5,2) -- (5,1.8) node[right] {$s$};
\draw[black] (5,1.8) -- (5,1.6) {};
\draw[black] (5,1.6) -- (5.4,1.6) {};
\draw (3,0) node[crossr] {};
\draw (4,0) node[crossp] {};
\draw[black]  (3.8,0.1) -- (3.8,-0.1) node[midway,below=0.18cm,align=center] {$\ M^2$};
\draw (4.5,0) node[crossp] {};
\draw (5,0) node[crossp] {};
\draw (5.5,0) node[crossp] {};
\end{scope}
\end{tikzpicture}
     \caption{
  $M_u^{I=2}(s,-s-t)$ 
    % pole structure for $u<0.\qquad\qquad\qquad$
  }
    \label{fig:utchannel}
  \end{subfigure}
  \end{center}
  \caption{\it Analytic  structure of  $M_t^{I=2}$ at fixed $u<0$ (a), fixed $t<0$ (b) and of $M_u^{I=2}$ at fixed $t<0$ (c). We denote by $C_0$,  $C_\infty$ (to be taken at $|s|\to \infty$) and the discontinuity along the real axis  the relevant contours of integration used for  the  dispersion relations.}
    \label{fig:sumrules}
\end{figure}

\subsubsection{Elastic Process}
For the  elastic process the dispersion relations that we will consider involve  $M^{I=2}_t$ at fixed $u<0$ and fixed $t<0$, and  $M^{I=2}_u$ at fixed $t<0$ (all in the $s-$plane). The analytic structure of these amplitudes and the contours we choose for the dispersion relations can be found in Fig.~\ref{fig:sumrules}. For the elastic process we define the high-energy average along the lines of \cite{Caron-Huot:2020cmc} as
\be
\big\langle (...) \big\rangle_{+-}^\pm \equiv\frac{1}{\pi}\sum_i (2J+1)  \int_{M^2}^\infty  \frac{dm^2}{m^2}\rho^{+-,\pm}_J(m^2) (...)\ .
\ee
\begin{itemize}
\item \textbf{$M_t^{I=2}\ u-$fixed:}
\end{itemize}
The integral of $M_t^{I=2}(s,u)/s^{k+1}$ along the contour $C_\infty$ of Fig.~\ref{fig:uchannel} vanishes for $k\geq{k_{\rm min}}$, due to \eq{UVa}. Because of the amplitude's analyticity, we can deform $C_\infty$ into the blue contour in Fig.~\ref{fig:uchannel} giving
\be
\oint_{\rm C_0} ds'\ \frac{M_t^{I=2}(s',u)}{s^{\prime k+1}}= 2 i
\int_{M^2}^\infty  ds'\ \frac{{\rm Im} M_t^{I=2}(s',u)}{s^{\prime k+1}}\,.
\label{SU}
\ee
Assuming $k_{\rm min}=1$,
we are able to derive, inserting \eq{pwpm} into \eq{SU} and expanding for $u\rightarrow0$, the following system of relations:
\bea
(k=1)\quad g_{0,0} u +g_{1,0} u^2 + ... &=&  \left\langle \frac{(-1)^J {\cal J}^2}{2m^4}u+\frac{(-1)^J {\cal J}^4}{6m^6}u^2+...\right\rangle_{+-}^-\,, \nonumber\\
(k=2)\quad  g_{1,0} u + g_{2,1} u^2 + ... &=&  \left\langle \frac{(-1)^J {\cal J}^2}{2m^6}u+\frac{(-1)^J {\cal J}^4}{6m^8}u^2+...\right\rangle_{+-}^- \ , \nonumber\\
&\vdots& 
\label{systemsu}
\eea
where ${\cal J}^2= J(J+1)$, ${\cal J}^4= (J-1)J(J+1)(J+2)$, etc... . The system above allows us to relate the low-energy Wilson coefficients to the high-energy UV-averages:
\be
g_{0,0}=   \left\langle \frac{(-1)^J {\cal J}^2}{2m^4}\right\rangle_{+-}^- \quad , \quad g_{1,0}=  \left\langle \frac{(-1)^J {\cal J}^2}{2m^6}\right\rangle_{+-}^- \quad , \ ... \ .
\label{wilsonselastic}
\ee
Furthermore one can notice that the system is over-constrained and therefore it is possible to find a set of non trivial null constraints, the first of which  starts at  $\mathcal{O}(1/m^6)$:
\be
\left\langle \frac{(-1)^J {\cal J}^2}{2m^6}\right\rangle_{+-}^- = \left\langle \frac{(-1)^J {\cal J}^4}{6m^6}\right\rangle_{+-}^-.
\ee
Nevertheless, as we will show later,  the null constraints  needed to bound the chiral anomaly are only those at  $\mathcal{O}(1/m^4)$.%For this reason we will not go into the details of deriving a general expression for the $u-$fixed null constraints, as for example done in \cite{}.

\begin{itemize}
\item \textbf{$M_t^{I=2}\ t-$fixed:}
\end{itemize}
We can repeat the arguments explained above for $M^{I=2}_t(s,-s-t)$ at fixed $t<0$. The integral identity one gets using Cauchy's theorem, following Fig.~\ref{fig:tchannel}, is
\be
\frac{1}{2 i}\oint_{\rm C_0} ds'\ \frac{M_t^{I=2} (s',-t-s')}{s^{\prime k+1}}=
\int_{M^2}^\infty  ds'\ \frac{{\rm Im} M_t^{I=2}(s',-t-s')}{s^{\prime k+1}}
+(-1)^k\int_{M^2}^\infty  ds'\ \frac{{\rm Im} M_t^{I=2}(s',-t-s')}{(s'+t)^{k+1}}\,. 
\label{ST}
\ee
This dispersion relation leads to a system of equations \footnote{We can assume here $k_{\rm min}=2$ as $k=1$ does not provide additional information.}
\bea
(k=2) \quad -g_{0,0} +g_{1,0} t + ... &=&  \left\langle \frac{2}{m^4}\right\rangle_{+-}^- + \left\langle\frac{2 {\cal J}^2-5}{m^6}\right\rangle_{+-}^- t + ... \,,\label{g00}\\
&\vdots&\nonumber
\eea
that is not over-constrained. Therefore,  by themselves, these equations do not lead to  any null constraint. Neverthless, we can combine it with the $u-$fixed system in \eq{systemsu} to get a new set of null constraints. 
We will be interested in the one at $\mathcal{O}(1/m^4)$ which we will need to constrain the anomaly. This is given by
\be
\left\langle \frac{(-1)^J {\cal J}^2 +4}{2m^4}\right\rangle_{+-}^- = 0	\,.
\label{ncelastic}
\ee
Notice that the averages are taken with  $R_{I=0}^{+-}-R_{I=1}^{+-}$, and therefore positivity is not guaranteed. 

\begin{itemize}
\item \textbf{$M_u^{I=2}\ t-$fixed:}
\end{itemize}
The amplitude $M_u^{I=2}=A_t(s,u)+\tilde{A}(s,u)$ at  $t-$fixed is the only one whose residues are all positive,
as can be seen from Table~\ref{tab:tablecouplings}.  Its has poles only in the  $s-$channel, 
as can be seen in Fig.~\ref{fig:utchannel}. 
According to \eq{pwpm} the high-energy average will now have a positive spectral density 
 $\rho_J^{+-,+}(s)>0$.
%\bea
%{\rm Im} M_u^{I=2}(s,u)\big|_{+-} &=& \sum_i (2J+1)\rho^{+-,u}_J(s) d_{1,1}^J\big(\cos \theta_s \big)\\
%(2J+1)\rho^{+-,t}_J(s) &=&\pi\sum_i\bigg(\frac{R_{I=0}^{+-}}{3}+\frac{R_{I=1}^{+-}}{2}\bigg)_i m_i^2 \delta(s-m_i^2) \delta_{J J_i},\nonumber
%\label{pwpmt}
%\eea
The dispersion relations look identical to \eq{SU}, but with the low-energy amplitude  given by \eq{elasticu}. Following the steps explained in the previous sections, we can find sum rules for the Wilson coefficients
of \eq{elasticu}, as for example (for $k=2$),
\be
h_{0,0}^s = \left\langle\frac{1}{m^4}\right\rangle_{+-}^+\,.
\label{h00}
\ee
The conditions  one gets from these dispersion relations (analogous to \eq{systemsu}) is however not over-constrained,
and therefore no additional null constraints are obtained. 
%All the null constraints at our disposal  do not have any notion of positivity of the spectral density. 

\subsubsection{Inelastic Process}
Let us   now focus on the inelastic process.  In particular, out of the three possible dispersion relations listed in Fig.~\ref{fig:sumrules}, we will only consider  $M_t^{I=2}$ at fixed $u<0$ and  fixed $t<0$ in the  $s-$plane.\footnote{ We will not consider $M_u^{I=2}$ in the inelastic case.  The corresponding dispersion relations were only  useful for the elastic case to relate     Wilson coefficients, such as $h^s_{0,0}$,  to sums over positive residues.}  
   We can proceed similarly as in the previous sections, but with the  low-energy amplitude  given  in \eq{inelastict}, the partial-wave decomposition  given in \eq{pwpp}, and 
the high-energy average defined by
\be
\big\langle (...) \big\rangle_{++}^\pm \equiv\frac{1}{\pi}\sum_i (2J+1)  \int_{M^2}^\infty  \frac{dm^2}{m^2}\rho^{++,\pm}_J(m^2) (...)\ .
\ee
\begin{itemize}
\item \textbf{$M_t^{I=2}\ u-$fixed:}
\end{itemize}
Once again the integral of $M_t^{I=2}(s,u)/s^{k+1}$ along the contour $C_\infty$ of Fig.~\ref{fig:uchannel} vanishes for $k\geq{k_{\rm min}}$, due to \eq{UVa}. We have again  \eq{SU} that  expanded  for small $u$ gives
\bea
(k=1)\quad\quad\quad\ \ \  f_{0,0}+2f_{1,0} u + ... &=&  \left\langle \frac{(-1)^{J+1} }{m^2}-\frac{(-1)^J (J^2+J-1)}{m^4}u+...\right\rangle_{++}^- \,,\nonumber \\
(k=2)\quad  f_{1,0} +(f_{2,1} +f_{2,0} )u + ... &=&  \left\langle \frac{(-1)^{J+1}}{m^4}-\frac{(-1)^J (J^2+J-1)}{m^6}u+...\right\rangle_{++}^-\ , \nonumber\\
&\vdots& 
\label{systemsuin}
\eea
leading to a new set of null constraints. At $\mathcal{O}(1/m^4)$, we have
\be
\left\langle \frac{(-1)^J({\cal J}^2 -3)}{2m^4}\right\rangle_{++}^- = 0 \,.
\label{ncinelastic}
\ee
%($\mathcal{O}(1/m^4)$ and therefore relevant for the chiral anomaly bound as we will discuss later on). 
\begin{itemize}
\item \textbf{$M_t^{I=2}\ t-$fixed:}
\end{itemize}
For fixed $t<0$, we can obtain new dispersion relations that  combined with the system \eq{systemsuin} leads to a second set of null constraints. By doing this, one can  notice that the first new null constraint enters at order $\mathcal{O}(1/m^6)$.

\subsubsection{Bounds on  Wilson coefficients}
From the above dispersion relations it is already possible to obtain  interesting bounds on the low-energy Wilson coefficients of the $W\pi W\pi$ amplitude, that are related to physical quantities such as the  dipole and quadrupole polarizabilities of the pions (see Appendix~\ref{pola}). A more detailed discussion is given in  Appendix~\ref{wilsonsapp}.
For example,  at ${\cal O}(s^2)$, we find the bounds 
\be
-2\leq \frac{g_{0,0}}{h^s_{0,0}} \leq 2\,, \ \ \ \ -1\leq \frac{f_{1,0}}{h^s_{0,0}} \leq 1\,.
\label{g00h00bound}
\ee
For higher-oder  Wilson coefficients a  numerical analysis is sometimes needed in order to find  the allowed values. In Figure~\ref{fig:g2g2p} we provide an example (see  Appendix~\ref{wilsonsapp} for details).

\begin{figure}[t]
\begin{center}
\includegraphics[width=0.5\linewidth]{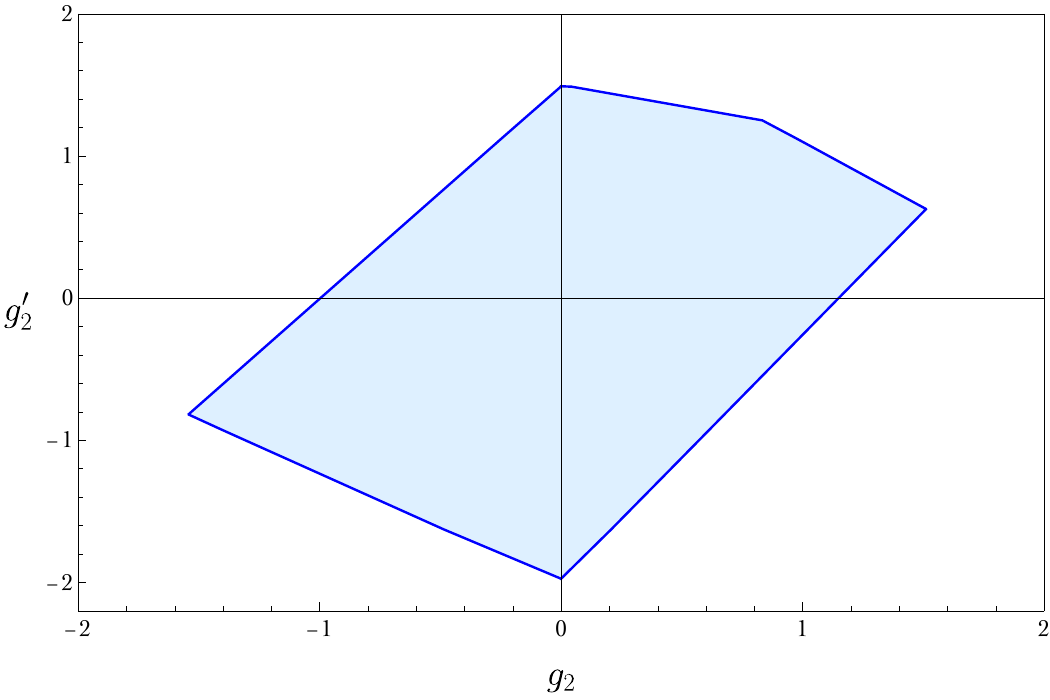}
\caption{\it Exclusion plot for the Wilsons $g_2$ and $g_2'$ given in \eq{g2g2pdef}.}
\label{fig:g2g2p}
\end{center}
\end{figure}

\subsection{The $W \pi\to \eta \pi$ Amplitude and the Chiral Anomaly}
Let us now consider  the   $W \pi\to\eta \pi$ amplitude. For  $\lambda_W=+1$  helicity, we have 
\be 
\mathcal{M}(W^{a+} , \pi^b,\eta, \pi^c) \propto  [12]\langle 24 \rangle [41]\propto \sqrt{stu}\,.  
\ee   
At low-energy, this amplitude can be written as
\be \label{eq:Anomaly}
|\mathcal{M}(W^{a+} ,\pi^b, \eta, \pi^c) |=
|f_{abc}|\mathcal{M}_{W \pi\eta \pi}
=\frac{\kappa}{2\sqrt{2}}  |f_{abc}| \sqrt{stu}+...\,.  
\ee   
The Wilson coefficient $\kappa$ is related to the chiral anomaly. Indeed, from the WZW term ~\cite{WESS197195,WITTEN1983422} we have 
\be
 -\frac{N_c}{48\pi^2} \varepsilon^{\mu \nu \alpha \beta} \Tr[A_{\mu L} U_{\nu L} U_{\alpha L} U_{\beta L} +\text{L} \to \text{R}]   \subset\mathcal{L}_{WZW}
 \label{WZW}\,,   
\ee
where $U_{\nu L} =(\partial_\nu U) U^\dagger$, $U_{\nu R} =U^\dagger(\partial_\nu U) $, and  $U=
\text{Exp}(i \eta /F_\pi) \text{Exp}(2i\pi^a \tau^a/F_\pi)$  with $\tau^a =\sigma^a/2$ and $A^a_{L}=A^a_R=W^a$.
By matching \eq{eq:Anomaly} with \eq{WZW}, we obtain
\be
\kappa = \frac{N_c}{12\pi^2 F_\pi^{3}}\,.
\label{kappa}
\ee

The $W \pi \eta \pi$ amplitude  is mediated by the following meson states:
\begin{equation}
\begin{tikzpicture}[baseline=.1ex]
  \node[text width=7cm] at (-0.7,0) {$s/u-\text{channel:}$};
  \begin{feynman}[every dot={/tikz/fill=black!70}]
    \vertex (e) at (0,-0.5) {\scriptsize{$G=-1$}} ;
    \vertex (e) at (0,0.5) {\scriptsize{$I=1$}} ;
    \vertex[dot] (m1) at (-0.4, 0){};
    \vertex[dot] (m2) at (0.4, 0){};
    \vertex (a) at (-1,-1) {$\pi$} ;
    \vertex (b) at ( 1,-1) {$\pi$};
    \vertex (c) at (-1, 1) {\small{$W$}};
    \vertex (d) at ( 1, 1) {\small{$\eta$}};
    \diagram* {
      (a) -- [dashed] (m1) -- [photon] (c),
      (m2) -- [double] (m1),
      (b) -- [dashed] (m2) -- [dashed] (d),
    };
  \end{feynman}
\end{tikzpicture}
\begin{tikzpicture}[baseline=.1ex]
  \node[text width=7cm] at (-0.6,0) { $t-\text{channel:}$};
  \begin{feynman}[every dot={/tikz/fill=black!70}]
      \vertex (e) at (-0.7,0) {\scriptsize{$G=1$}} ;
    \vertex (e) at (0.7,0) {\scriptsize{$I=1$}} ;
    \vertex[dot] (m1) at (0,0.4){};
    \vertex[dot] (m2) at (0,-0.4){};
    \vertex (a) at (-1,-1) {$\pi$} ;
    \vertex (b) at ( 1,-1) {$\pi$};
    \vertex (c) at (-1, 1) {\small{$W$}};
    \vertex (d) at ( 1, 1) {\small{$\eta$}};
    \diagram* {
      (d) -- [dashed] (m1) -- [photon] (c),
      (m2) -- [double] (m1),
      (b) -- [dashed] (m2) -- [dashed] (a),
    };
  \end{feynman}
\end{tikzpicture}
\label{anomalystatesplot}
\end{equation}
For the $s/u-$channel the exchanged mesons must  have $G=-1$,  $I=1$,  $J=\text{even}$ and parity $P=(-1)^J=+1$. These states are classified as $n=2$ in Table.~\ref{tab:states} (highlighted in green).
On the other hand,  in the $t-$channel the states must have  $G=+1$,  $I=1$,  $J=\text{odd}$ and  $P=(-1)^J=-1$. 
These states are classified as $n=5$ in Table.~\ref{tab:states} (also highlighted in green). 
We notice therefore that in this amplitude, only 2 out of the 6 possible types of mesons  contribute.

\subsection{Dispersion Relations}
%We first derive the dispersion relation of WZW anomaly through $W \eta \pi \pi$ amplitude. We can find that at lower energy limit, its amplitude in Eq.~(\ref{eq:Anomaly}) is not analytical due to its polarization structure, 
%To overcome this issue, we can first divied the 

Let us consider the $W\pi\eta\pi$ amplitude  without the non-analytical pre-factor coming from  the  polarization structure,
$\mathcal{M}_{W\pi\eta \pi}/\sqrt{stu}$, and assume at  fixed $u<0$ 
\be
\lim_{|s| \to \infty} \frac{\mathcal{M}_{W\pi\eta \pi}}{\sqrt{stu}}= 0\,,
\ee
and similarly at fixed $t<0$.
In the $s$-plane with $u$ fixed, we get the following dispersion relation:
 \be
 \oint \frac{ds}{2\pi i} \frac{\mathcal{M}_{W\pi\eta \pi}(s,u)}{\sqrt{-s(s+u) u}} 
 =\sum_{i}^{J^+_\text{even}}  \frac{ g_{W\pi i} g_{\pi \eta i}
 \, d^J_{1,0}(-1-\frac{2u}{m_i^2}) }
 {\sqrt{-m_i^2 (m_i^2+u)u}}  +
\sum_{i}^{J^-_\text{odd}}
  \frac{g_{W\eta i} g_{\pi \pi i}\, d^J_{1,0}(-1-\frac{2u}{m_i^2}) }
 {\sqrt{-m_i^2 (m_i^2+u)u}} \,,
\ee
where we specified that only $J^P$ mesons with $J$ odd or even are summed over.
Expanding the above dispersion relation for $u \to 0$, we obtain 
\bea \label{eq:ano1}
\frac{\kappa}{2\sqrt{2}} =\sum_{i}^{J^-_\text{odd}}  \frac{g_{W\eta i} g_{\pi \pi i}\, {\cal J}}{m_i^{3} }
-  
\sum_{i}^{J^+_\text{even}}
 \frac{ g_{W\pi i} g_{\pi \eta i} \, {\cal J} }{m_i^{3} }   \,,
\eea
where ${\cal J}=\sqrt{{\cal J}^2}$.

Similarly, we can also get  another dispersion relation by fixing $t<0$ in the $s$-plane. 
%Again, to project out $\kappa$, the high-energy behavior of the amplitude $\mathcal{M}^\prime$ should satisfy the constraint $\lim_{t \to \infty} M^\prime_{W\eta \pi \pi} (s,t) /t \to 0$ with $s$ fixed, 
%\bea 
%k=1: \quad  \oint \frac{dt}{2\pi i} \frac{\mathcal{M}_{W^a \eta \pi^b \pi^c}(s,t)}{t^{2}\sqrt{s(1+\frac{s}{t})}}=\sum_{J=\text{even}}  \frac{g_{2}^J g^{J \ast}_{\pi \eta } d^J_{1,0}(1+\frac{2s}{m_i^2}) }{m_i^{2}\sqrt{s(1+\frac{s}{m_i^2})}} +\sum_{J=\text{even}}  \frac{g_{2}^J g^{J \ast}_{\pi \eta } m_i^2 d^J_{1,0}(1+\frac{2s}{m_i^2}) }{(m_i^{2} +s)^{2}\sqrt{s(\frac{m_i^2}{m_i^2 +s})}} \,.
%\eea 
%Expanding the above dispersion relation for $s \to 0$, we can get another dispersion relation of $\kappa$,
We obtain in this case
\be
\label{eq:ano2}
\frac{\kappa}{2\sqrt2} = 2 
\sum_{i}^{J^+_\text{even}}
 \frac{ g_{W\pi i} g_{\pi \eta i} \, {\cal J} }{m_i^{3} }   \,.
\ee
From \eq{eq:ano1} and \eq{eq:ano2} we get the null constraint
\be
 \sum_{i}^{J^-_\text{odd}}  \frac{g_{W\eta i} g_{\pi \pi i}\, {\cal J}}{m_i^{3} } 
=3 
\sum_{i}^{J^+_\text{even}}
 \frac{ g_{W\pi i} g_{\pi \eta i} \, {\cal J} }{m_i^{3} }\,.
 \label{supernc}
\ee
At large $N_c$,   the coupings of singlet $\eta$ can be related with the $\pi^a$ couplings:
\bea 
g_{W\eta i} = (g_{5})_i\  \  \ \ (J^-_\text{odd}) \quad , \quad g_{\pi \eta  i} =g_{\pi \pi i}\ \  \ \ (J^+_\text{even})\,. 
\eea 
Using these relations, we can express  $\kappa$ and the null constraint \eq{supernc} as
\bea
\frac{\kappa}{4\sqrt{2}}&=&\sum_{i}^{J^+_\text{even}}
\frac{(g_{2}g_{\pi\pi})_{i}\mathcal{J}}{m_{i}^3}\label{anomalync0} \\
 &=& 
  \sum_{i}^{J^-_\text{odd}}
 \frac{(g_{5}g_{\pi\pi})_{i}\mathcal{J}}{3m_{i}^3} \ .
\label{anomalync}
\eea

\section{Bounding the Chiral Anomaly}
\label{sec:bound}
In this section we analytically derive an upper bound on the anomaly coefficient $\kappa$ using
 the $W\pi\to W\pi$ null constraints    \eq{ncelastic} and \eq{ncinelastic}.
These two null constraints are enough to get a bound on $\kappa$.
We will however improve the bound by  incorporating  
the null constraint \eq{supernc} from $W\pi\to \eta\pi$.

Using the identity (for $a_i,b_i\in \mathbb{R}$)
\be
\sum_i a_i b_i \leq \sqrt{\sum_i a_i^2 b_i^2}\leq \sqrt{\big(\sum_i a_i^2 \big) \big(\sum_i b_i^2\big)}\ ,
\ee
we can obtain a bound on $\kappa$  using \eq{anomalync0} and \eq{anomalync} respectively
\bea
\frac{\kappa}{4\sqrt{2}} &\leq& \sqrt{\bigg( \sum^{J^+_\text{even}}_i \frac{g_{\pi\pi i}^2}{m_{i}^2}\bigg) \bigg\langle \frac{\mathcal{J}^2}{m^4} \bigg\rangle_2}\equiv\kappa_2^\text{UB}\,, \nonumber\\
\frac{\kappa}{4\sqrt{2}} &\leq& \frac{1}{3}\sqrt{\bigg( \sum^{J^-_\text{odd}}_i \frac{g_{\pi\pi i}^2}{m_{i}^2} \bigg) \bigg\langle \frac{\mathcal{J}^2}{m^4} \bigg\rangle_5}\equiv\kappa_5^\text{UB}\,. 
\label{ubanomaly2}
\eea
where we have defined
\be
\langle (...) \rangle_n \equiv\frac{1}{\pi}\sum_i (2J+1)  \int_{M^2}^\infty  \frac{dm^2}{m^2}\rho^{n}_J(m^2) (...)\,,
\label{hean}
\ee
with
\be
(2J+1)\rho^{n}_J(m^2) =\pi\sum_i(g_{n}^2)_i m_i^2 \delta(m^2-m_i^2) \delta_{J J_i}\ ,
\ee
and $g_{n}^2\geq0$ defined in \eq{defgn}. 
The bounds \eq{ubanomaly2} depend on ${\cal J}^2/m^4$ high-energy averages which  
can be written as a function of $1/m^4$ averages by using the null constraints
\eq{ncelastic} and \eq{ncinelastic}, rewritten as
\bea
+\bigg\langle \frac{\mathcal{J}^2-4}{m^4}\bigg\rangle_1-\bigg\langle \frac{\mathcal{J}^2+4}{m^4}\bigg\rangle_2-\bigg\langle \frac{\mathcal{J}^2+4}{m^4}\bigg\rangle_3-\bigg\langle \frac{\mathcal{J}^2-4}{m^4}\bigg\rangle_4-\bigg\langle \frac{\mathcal{J}^2-4}{m^4}\bigg\rangle_5+\bigg\langle \frac{\mathcal{J}^2+4}{m^4}\bigg\rangle_6=0\,, \ \nonumber \\
-\bigg\langle \frac{\mathcal{J}^2-3}{m^4}\bigg\rangle_1-\bigg\langle \frac{\mathcal{J}^2-3}{m^4}\bigg\rangle_2+\bigg\langle \frac{\mathcal{J}^2-3}{m^4}\bigg\rangle_3+\bigg\langle \frac{\mathcal{J}^2-3}{m^4}\bigg\rangle_4-\bigg\langle \frac{\mathcal{J}^2-3}{m^4}\bigg\rangle_5-\bigg\langle \frac{\mathcal{J}^2-3}{m^4}\bigg\rangle_6=0\,.\ 
\label{nc123456}
\eea
By summing these two relations we get
\be
\bigg\langle \frac{2\mathcal{J}^2}{m^4}\bigg\rangle_2+\bigg\langle \frac{2\mathcal{J}^2}{m^4}\bigg\rangle_5= -\bigg\langle \frac{1}{m^4}\bigg\rangle_1-\bigg\langle \frac{1}{m^4}\bigg\rangle_2-\bigg\langle \frac{7}{m^4}\bigg\rangle_3+\bigg\langle \frac{1}{m^4}\bigg\rangle_4+\bigg\langle \frac{7}{m^4}\bigg\rangle_5+\bigg\langle \frac{7}{m^4}\bigg\rangle_6\ .
\label{ncpppm}
\ee
\eq{ncpppm}   gives an  interesting relation between the two  $\mathcal{J}^2/m^4$ averages  involved in \eq{ubanomaly2} that  allows to find the following relation
\be
\bigg\langle \frac{\mathcal{J}^2}{m^4}\bigg\rangle_2+\bigg\langle \frac{\mathcal{J}^2}{m^4}\bigg\rangle_5= 3 f_{1,0}
-2g_{0,0}\equiv {\cal P}\geq 0\,,
\label{h00k}
\ee
where we have used 
\bea
g_{0,0} &=&  \bigg\langle \frac{2}{m^4}\bigg\rangle_1+\bigg\langle \frac{2}{m^4}\bigg\rangle_2+\bigg\langle \frac{2}{m^4}\bigg\rangle_3-\bigg\langle \frac{2}{m^4}\bigg\rangle_4-\bigg\langle \frac{2}{m^4}\bigg\rangle_5-\bigg\langle \frac{2}{m^4}\bigg\rangle_6\,, \label{g002} \\
 f_{1,0} &=&  \bigg\langle \frac{1}{m^4}\bigg\rangle_1+\bigg\langle \frac{1}{m^4}\bigg\rangle_2-\bigg\langle \frac{1}{m^4}\bigg\rangle_3-\bigg\langle \frac{1}{m^4}\bigg\rangle_4+\bigg\langle \frac{1}{m^4}\bigg\rangle_5+\bigg\langle \frac{1}{m^4}\bigg\rangle_6\, ,\label{f102} 
\eea 
obtained from the  sum rules  \eq{g00} and \eq{systemsuin} respectively.

We can also obtain a relation between the sum over $g^2_{\pi\pi i}/m_i^2$ in \eq{ubanomaly2} by 
using sum rules derived from dispersion relations for the   $\pi \pi \rightarrow \pi \pi$ process in \cite{Albert:2022oes,Fernandez:2022kzi}. 
In particular, we need 
\be
F_\pi^{-2}= 2 \bigg( \sum^{J^+_\text{even}}_i\frac{g_{\pi\pi i}^2}{m_{i}^2}+ \sum^{J^-_\text{odd}}_i\frac{g_{\pi\pi i}^2}{m_{i}^2}\bigg)\,. 
\label{fpi}
\ee

Armed with   \eq{h00k} and \eq{fpi}, we can now obtain a bound on $\kappa$ as a function of $F^2_\pi$ and 
$\mathcal{P}$.
Defining 
\be
X\equiv F_\pi^{2} \sum^{J^-_\text{odd}}_i \frac{g_{\pi\pi i}^2}{m_{i}^2} \quad \text{and} \quad  Y\equiv \frac{1}{\mathcal{P}}\bigg\langle \frac{\mathcal{J}^2}{m^4} \bigg\rangle_5\ ,
\ee
we can rewrite $\kappa_5^\text{UB}$ and $\kappa_2^\text{UB}$ as
\be
\frac{\kappa_5^\text{UB}}{\sqrt{\mathcal{P}/ F_\pi^{2}}} = \frac{1}{3}\sqrt{ X Y}\ ,\ \ \ \ 
\frac{\kappa_2^\text{UB}}{\sqrt{\mathcal{P}/ F_\pi^{2}}} \leq \sqrt{\bigg( \frac{1}{2}-X\bigg)\bigg(\frac{1}{2}-Y\bigg)}\,.
\label{ubanomaly3}
\ee
Written in this way we notice that as we decrease $X$ and $Y$ to make   $\kappa_2^\text{UB}$ larger, $\kappa_5^\text{UB}$ becomes smaller, and viceversa.
 Since $\kappa$ must be smaller than both bounds, the largest value of $\kappa$   is reached when $\kappa_2^\text{UB}=\kappa_5^\text{UB}$. This is achieved when
\be
\frac{1}{9} X Y=\left( \frac{1}{2}-X\right)\left(\frac{1}{2}-Y\right)  \quad \rightarrow \quad Y= \frac{X-1/2}{16X/9-1}\,,
\label{Ymax}
\ee
that inserted in \eq{ubanomaly3} gives us $\kappa_5^\text{UB}$  as a  function of only $X$ that is maximized  for  $X=3/8$.    This corresponds  to $Y=3/8$, and gives $\kappa_5^\text{UB}|_{\rm max}=1/8$. Taking this value 
in \eq{ubanomaly2}, we finally get the following upper bound on the chiral anomaly:
\be
\frac{\kappa}{\sqrt{\mathcal{P}/F_\pi^{2}}} \leq \frac{1}{\sqrt{2}}\,.
\label{anomalybound}
\ee
This is the main result of the paper. This tells us that the anomaly coefficient is bounded by $\mathcal{P}$,
a quantity related with the  polarizabilities of the pions (see Appendix~\ref{pola}).
Using the constraints  \eq{g00h00bound}, we can write \eq{anomalybound} as a function of other Wilson coefficients.
For example, we get that ${\cal P}\leq 7 h^s_{0,0}$ that leads to  
\be
\frac{\kappa}{\sqrt{h_{0,0}^s /F_\pi^{2}}} \leq \sqrt{\frac{7}{2}}\,.
\label{h00bound}
\ee
 
In the case of $SU(N_c)$ gauge theories with $N_f$ quarks, the anomaly coefficient $\kappa$ is known, and  \eq{anomalybound}  provides  a bound 
on the Wilson coefficients.
% (let us take $h^s_{0,0}$ for example using \eq{h00bound}).
For example, 
using \eq{kappa} and $F_\pi\simeq \sqrt{N_c/3} \, m_\rho/7$ \cite{Bali:2013kia}, we obtain from \eq{h00bound}
\be
h^s_{0,0}\gtrsim \frac{0.44}{m_\rho^4}\,.
\ee
As in \cite{Albert:2022oes,Fernandez:2022kzi},
it is also  interesting to obtain the predictions of  different  models to the ratio \eq{anomalybound},
in order to understand how close these are to the upper bound. 
For example, in models of only scalars,
we have that  $h^s_{0,0}$ is zero and then $\kappa=0$. 
For  $su$-models (introduced in \cite{Caron-Huot:2020cmc} for the first time, and studied in \cite{Caron-Huot:2021rmr,Albert:2022oes,Fernandez:2022kzi} for four-pion amplitudes), we show in  Appendix~\ref{sumodelapp} that
\be
\frac{\kappa}{\sqrt{{\cal P}/F_\pi^{2}}} \Bigg|_{su-\text{model}}\lesssim \frac{0.97}{\sqrt{2}}\,, 
\label{kappaboundsu}
\ee
which falls within the bound established in \eq{anomalybound}, only   $\sim 3\%$ away  from saturation.

 \eq{anomalybound} gives a non-trivial constraint on  phenomenological models for  QCD, such as NJL
 or  holographic models. 
 For example, in   holographic models  the chiral anomaly arises from a Chern-Simons (CS) term.
When the corresponding dual 4D theory is not known, the CS coefficient cannot be matched to the UV theory
and therefore cannot be determined.
\eq{anomalybound} provides in this case a bound on the the size of the CS term.
This could also be useful for  models of axions.

\section{Conclusions}
\label{sec:conclusion}
The analyticity, unitarity and the good  high-energy behaviour of
 scattering amplitudes  provide severe constraints on  the low-energy physical quantities of gauge theories like QCD.
 In this work we have studied scattering amplitudes 
  involving external gauge fields and goldstones ($\pi$ and $\eta$) in the  large-$N_c$ limit
  to obtain several new constraints. 
  
  Our main result has been to provide an analytical  bound on the chiral anomaly, \eq{anomalybound}.
To derive  this bound we have first analyzed the  $W\pi \rightarrow W\pi$ amplitude,
 providing     the selection rules for the  mesons exchanges, as well as the  sign of their on-shell residues. 
We have considered  the elastic ($W^+\pi \rightarrow W^+\pi$)
and inelastic ($W^+\pi \rightarrow W^-\pi$) processes, as both are needed to derive
 $\mathcal{O}(1/m^4)$ null constraints on the meson couplings.\footnote{We have found that adding more null constraints does not improve the bound.} 
%These null constraints have been crucial to obtain the bound  \eq{anomalybound}.
We have also derived sum rules for the Wilson coefficients of these amplitudes. Of especial use has been
the sum rule for $h^s_{0,0}$ that has shown to have all its terms positive.

We have later  considered the  $W\pi\to\eta\pi$ amplitude, which  at low energy yields the chiral anomaly coefficient $\kappa$. 
We have shown, once again via dispersion relations, that there are  two different ways to determine $\kappa$, 
implying  a  non-trivial  constraint among the corresponding meson couplings.

Putting all this together, we have been able to derive analytically  \eq{anomalybound}.
This bound has interesting implications for UV completions of these amplitudes, as we 
have shown for the case of large-$N_c$ QCD.
We have also  considered  $su$-models (see Appendix~\ref{sumodelapp}) that seem to almost saturate the bound.
We hope that in the future this  bound could  be tested in lattice simulations. 
Finally,  we have also briefly study the bounds on  other Wilson coefficients,  showing
the EFT-hedron  geometry \cite{Arkani-Hamed:2020blm}  of their allowed parameter space -see for example Fig.~\ref{fig:g2g2p}.
 
In the future,  it could 
also  be interesting to study in more detail which theories saturate these bounds. 
  Similarly, our analysis  can be extended to understand implications  on composite Higgs models  and bounds on dimension six operators such as    $H^\dagger H  F_{\mu \nu} F^{ \mu \nu}$.

\hskip1.1cm 

{\bf Note added:} While preparing this article, it was submitted to
 the archives  Ref.~\cite{Albert:2023jtd} that also uses dispersion relations
to  numerically obtain    bounds on the chiral anomaly coefficient.

\hskip.8cm

\section*{Acknowledgments}
We are very grateful to Francesco Riva and Pyungwon Ko for their valuable insights.
This work   has been  partly supported by the research grants 2021-SGR-00649 and PID2020-115845GB-I00/AEI/10.13039/501100011033.

\appendix
\section{$q\bar q$ mesons and the sign of  on-shell residues}

\label{appa}
The $q\bar q$ mesons can be classified according to their quantum numbers: mass $m$, spin $J$, parity $P$,
Isospin $I$, and $G$-parity. 
Defining by $\ell$ the orbital angular momentum of the $q\bar q$ system, $s=0,1$ the total spin,  and $I=0,1$ the isospin, we have
\bea
J&=&\ell+s,...,\ell-s\,,\nonumber\\
P&=&(-1)^{\ell+1}\,,\nonumber\\
G&=&(-1)^{I+\ell+s}\,.
\eea
By taking  $\ell=0,1,2,...$, we can build the meson list of  Table~\ref{tab:states}. Notice that we have only six types of mesons that we label with $n=1,...,6$.

To understand the sign of the residues in \eq{Rstates}
arising from  meson exchanges in the process $W\pi \to W\pi$, we  proceed in the following way 
(recalling that the couplings $g_{W\pi i}$ are real):
\begin{itemize}
\item
For the elastic case  $W^+\pi \to W^+\pi$ (amplitude ${\cal M}^{+-}$), we obviously have that these are proportional to  
$g_{W\pi i}^2>0$ so the sign is always positive.  
\item
For  $W^+\pi \to W^-\pi$ (amplitude ${\cal M}^{++}$)  we can consider the process 
$W^+\pi \to R_i\to W^-\pi$ in the forward limit $t\to0$.
By performing a  $P$ transformation and a spacial rotation of 180$^0$ degrees,
the coupling in $W^+\pi \to R_i$  can be related to that of $R_i\to W^-\pi$.
Using that  $d_{1,0}^J (-\cos\theta) =(-1)^J d_{1,0}^J (\cos\theta)$,
we have then that $W^+\pi \to R_i\to W^-\pi$  is proportional to 
$g_{W\pi i}^2\times (-1)^{J_i}\times P_i$.
\end{itemize}
This leads to  the signs in \eq{Rstates} for the different type of mesons.

\section{Polarizabilities of the pions}
\label{pola}

Given the $\gamma\gamma\to\pi\pi$  scattering amplitude  with the pion exchange
subtracted, $ {\cal M}(\gamma^+,\gamma^\pm,\pi,\pi)$, the pion polarizabilities are  defined as 
the coefficients coming from an expansion in $s$ at fixed $t=m^2_\pi$
\cite{Guiasu:1979sz,Gasser:2006qa,Moinester:2022tba}:
\be
\frac{\alpha}{m_\pi} {\cal M}(\gamma^+,\gamma^\pm,\pi,\pi) (s,t=m^2_\pi)=(\alpha_1\mp \beta_1)_\pi
+\frac{s}{12} (\alpha_2\mp \beta_2)_\pi+...\,.
\label{poladef}
\ee
In particular,  $(\alpha_{1}\mp \beta_{1})_\pi$ and $(\alpha_{2}\mp \beta_{2})_\pi$ are respectively the dipole  and quadrupole polarizabilities of the pions $\pi=\pi^\pm,\pi^0$. 
These quantities have been measured experimentally \cite{Guiasu:1979sz,Gasser:2006qa,Moinester:2022tba} and also analyzed in lattice simulations
\cite{Feng:2022rkr}.
In our convention the photon corresponds to the gauging of  $Q=T_3+B/2$, 
a subgroup of the global $U(2)$, and we must
change $s\leftrightarrow t$ in  \eq{poladef}.
The dipole  polarizabilities receive then contributions from our $g_{0,0}$ and  $h^s_{0,0}$ for the $+-$ helicities of the photon (although they are suppressed by $m^2_\pi$), and $f_{0,0}$  for the $++$.
On the other hand, $f_{1,0}$ contributes to  the quadrupole polarizability.

\section{Bounds on the Elastic $W\pi W\pi$ Wilson coefficients}
 \label{wilsonsapp}
Let us  show here how to bound the parameter space of the Wilson coefficients appearing in \eq{elastict} and \eq{elasticu}.
 We are considering the elastic process since better bounds can be obtained in this case due to  positivity. 
Nevertheless, the same approach can also be used for   the inelastic case.

Let us begin with the simplest analytical bound arising from the sum rules of $g_{0,0}$ and $h^s_{0,0}$, 
\eq{g00} and \eq{h00} respectively, that gives
\be
\frac{g_{0,0}}{h^s_{0,0}} = -\frac{\left\langle \frac{2}{m^4}\right\rangle_{+-}^-}{\left\langle\frac{1}{m^4}\right\rangle_{+-}^+ }\,.
\ee
The absolute value of this ratio  must always be smaller or equal to 2, since the states that enter in $g_{0,0}$ and in $h^s_{0,0}$ are the same, the only difference being that in $h^s_{0,0}$ everything enters additively, while some of these states enter with a minus sign in $g_{0,0}$. 
The bound will be therefore  maximized when the only states exchanged are those who contribute negatively to $g_{0,0}$ (namely states with $n=1,2,3$), while it is  minimized when the states that contribute positively are the only ones being exchanged ($n= 4,5,6$).  This leads to the bound \eq{g00h00bound}, and similarly for $f_{1,0}$ from its sum rule in \eq{systemsuin}.

The same result can be found numerically  using \texttt{SDPB} \cite{Simmons-Duffin:2015qma}. To find this bound we must initially redefine the Wilson coefficients and null constraints (of both the elastic and inelastic processes) in terms of the high-energy average defined \eq{hean}, similarly to what was done in \eq{nc123456}. We can then construct the bootstrap equation 
\be
h_{0,0}\vec{v}_1+g_{0,0}\vec{v}_2+ \sum_{n=1..6}\langle \vec{v}_n(m^2,J)\rangle_n=0\,,
\ee
which holds true if $\vec{v}_1=(1,0,0,..), \vec{v}_2 = (0,1,0,...)$ and for an appropriate choice of the vectors 
$\vec{v}_n(m^2,J)$ (which contain the null constraints). Following canonical the optimization procedure explained in detail in \cite{Caron-Huot:2020cmc,Caron-Huot:2021rmr,Albert:2022oes}, one can bound numerically the Wilson coefficients. In particular, this method can be replicated for higher-order Wilson coefficients. For example, defining 
\be
g_2=\frac{g_{2,0}}{h_{2,0}^a+2h_{2,0}^s+h_{2,1}^s/2} \qquad \text{and}\qquad g_2'=\frac{g_{2,1}}{h_{2,0}^a+2h_{2,0}^s+h_{2,1}^s/2}\,, 
\label{g2g2pdef}
\ee
where the normalization has been chosen opportunely for positivity arguments as
\be
h_{2,0}^a+2h_{2,0}^s+h_{2,1}^s/2=\bigg\langle \frac{(\mathcal{J}^2-8)\mathcal{J}^2+14}{4m^8}\bigg\rangle_{+-}^+\, ,
\ee
we can repeat the numerical procedure and find the exclusion plot shown in Figure~\ref{fig:g2g2p}.

\section{The $su$-model}
\label{sumodelapp}

The $su$-models are defined as those leading to healthy   amplitudes with a single mass scale $m$. They  often predict Wilson coefficients at the boundaries of the allowed regions \cite{Caron-Huot:2020cmc,Caron-Huot:2021rmr,Albert:2022oes,Fernandez:2022kzi}.
For the $\pi\pi\to \pi\pi$ process the $su$-amplitude was already presented in \cite{Fernandez:2022kzi} 
where it  was shown to  take the general form 
\be
\mathcal{M}(s,u)\bigg|_{4\pi}=\frac{m^2(s+u)+\lambda su}{(s-m^2)(u-m^2)}\,,
\label{m4pi}
\ee
with -$2\leq\lambda\leq\frac{2\ln 2-1}{1-\ln 2}$.
In the limiting case  $\lambda=\frac{2\ln 2-1}{1-\ln 2}$, the amplitude \eq{m4pi} contains no poles associated to  $J=0$ states \cite{Fernandez:2022kzi}. In this case,
the residues in the $s$-channel of \eq{m4pi} are given by   
\be
g_{\pi \pi i}^2\big|_{4\pi} = \{0.78, 0.18,0.04, ... \}\,,
\label{gpipicouplings}
\ee
and correspond to states of $J^P=1^-,2^+,...$  respectively ($n=5$ and $n=2$ of  states $G=1$ alternating).
The amplitude \eq{m4pi} predicts  the Wilson coefficient
\be
 F_{\pi}^{-2} =\frac{2}{m^2}\,. 
 \label{fpisu}
\ee
 Notice that we have absorbed  an overall factor in  \eq{m4pi}  into $F_\pi^{-2}$.

An $su$-amplitude for $M^{I=2}_{t} (s,u) \big|_{+\pm}$ can also be constructed  
following the conditions:
\begin{itemize}
\item{It must have a single mass scale $m$.}
\item{No $t-$channel poles.}
\item{It must be proportional to $t$ and $su$ for $++$ and $+-$ amplitudes respectively.}
\item{It must drop  as    $M^{I=2}_{t} (s,u) /s\rightarrow 0$ for $|s|\rightarrow \infty$ at $t-$fixed, and similarly for $u-$fixed.}
\end{itemize}
The most general amplitude following these criteria takes the form
\be
M^{I=2}_{t} (s,u) \big|_{++} =-  \frac{m^2\, t}{(s-m^2)(u-m^2)}\,,\ \ \ \
M^{I=2}_{t} (s,u) \big|_{+-} = - \alpha\frac{su}{(s-m^2)(u-m^2)}\,,
\label{supi2}
\ee
where $\alpha$ is a constant.
The residues of \eq{supi2} in the $s$-channel are
\be
R^{++}_{I=0}-R^{++}_{I=1}= \{0.82,-0.15,0.03, ... \}\,,\ \ \
R^{+-}_{I=0}-R^{+-}_{I=1}=\alpha\, \{0.58,-0.09,0.02, ... \}\,.
\label{ressu}
\ee
corresponding to $J=1,2,3,...$ states, where for each $J_{\rm odd}$ ($J_{\rm even}$) 
they can be of type $n=1,4,5$ ($n=2,3,6$)   of Table~\ref{tab:states}.

From \eq{supi2} we obtain
\be
f_{1,0}=\frac{1}{m^4}\,, \ \ \ 
g_{1,0}=-\frac{\alpha}{m^4}\ \ \ \to \ \  {\cal P}=\frac{3+2\alpha}{m^4}\,.
\label{psu}
\ee
Notice that we have absorbed  an overall factor in  the $++$ amplitude in \eq{supi2} into $f_{1,0}$. 

According to \eq{Rstates}, we can obtain the couplings $g_{2}^2$ and $g_{5}^2$
(which are the ones that enter in the anomaly),  by adding the two sets of residues of \eq{ressu}. 
We obtain
\be
g_{W\pi i}^2\big|_{W^2\pi^2}=\{0.41+0.29\alpha, 0.07+0.05\alpha, ... \}\,,
\label{gWpicouplings}
\ee
corresponding to $n=5$ and $n=2$ states alternating. 
To have positive $g^2_{5,2}$, we must demand $-1.41\leq \alpha \leq 1.41$. 
The value $\alpha = 1.41\ (-1.41)$ maximizes (minimizes)  $g^2_{5,2}$.

Finally,  the $W\pi \eta \pi$ amplitude can be constructed with  the following conditions:
\begin{itemize}
\item Single mass scale $m$.
\item $s\leftrightarrow u$ crossing symmetric.
\item Proportional to $\sqrt{s t u}$. 
\item $s/u-$channel poles  associated  only to even-spin states (see \eq{anomalystatesplot}).
\item It must drop as ${\cal M}_{W\pi\eta\pi} /\sqrt{stu}\rightarrow 0$ for $|s|\rightarrow \infty$ at $t-$fixed, and similarly for $u-$fixed.
%$M_{W\pi\eta\pi} /t\rightarrow 0$ for $t\rightarrow \infty$ at $s-$fixed.
 \end{itemize}
The most general amplitude one can construct following these conditions is 
\be
{\cal M}_{W\pi\eta\pi}=-\beta m\sqrt{s t u}\, \frac{m^2/2+ t}{(s-m^2)(u-m^2)(t-m^2)}\ ,
\label{suanomaly}
\ee
where $\beta>0$ is a constant. We obtain the following residues in the $s$-channel:
\be
|g_{W\pi i}g_{\pi\pi i}|= \beta\, \{0.48,0.10,0.02,... \}\,,
\label{gWpipicouplings}
\ee
corresponding to  states  $n=5$ and $n=2$ of Table~\ref{tab:states} alternating.

From \eq{eq:Anomaly} and \eq{suanomaly} we get  the anomaly coefficient
\be
 \kappa=\frac{\sqrt{2}\beta}{m^3}\,.
 \label{kappasu}
 \ee
To maximize this value, we  must take the largest possible value of $\beta$.
Nevertheless, this is constrained by the fact that the couplings in \eq{gWpipicouplings} cannot overcome
\be
\sqrt{g_{W\pi i}^2\big|_{W^2\pi^2}\, g_{\pi\pi i}^2\big|_{4\pi}} = \{\sqrt{0.32+0.23 \alpha}, \sqrt{0.01+0.01 \alpha}, ... \}\,,
\label{product}
\ee
coming from  \eq{gpipicouplings} and \eq{gWpicouplings}, as there can always be more states $n=5,2$ (for a given $J$) in \eq{m4pi} and \eq{supi2}  than in \eq{suanomaly} (as this latter requires that the interchanged mesons must have both 
$g_{W\pi i}$ and $g_{\pi\pi i}$ nonzero). 
This  gives 
\be
\beta_{\rm max}=\sqrt{1.37+0.97\alpha}\,.
\label{bmax}
\ee

With \eq{fpisu}, \eq{psu}, \eq{kappasu} and  \eq{bmax}, one finds that $\kappa/\sqrt{{\cal P}/F^2_\pi}$  is maximized for $\alpha=1.41$, leading to \eq{kappaboundsu}.

\bibliography{biblio}
 \bibliographystyle{apsrev} 
 
\end{document}